\documentclass[preprint,authoryear,12pt]{elsarticle}

\usepackage{epsfig}
\usepackage{aas_macros}
\usepackage{amssymb}

\usepackage[ps2pdf,%
a4paper=true,%
breaklinks=true,%
colorlinks=true,%
pdfauthor={E. Bozzo, P. Romano, C. Ferrigno, P. Esposito, V. Mangano},%
pdftitle={Observations of Supergiant Fast X-ray Transients with LOFT}
]{hyperref}

\journal{Advances in Space Research}

\begin{document}

\begin{frontmatter}



\title{Observations of Supergiant Fast X-ray Transients with LOFT}


\author{E. Bozzo}
\address{ISDC, Data Center for Astrophysics of the University of Geneva, chemin d'Ecogia, 16 1290 Versoix Switzerland}
\ead{enrico.bozzo@unige.ch}

\author{P. Romano}
\address{INAF--IASF Palermo, Via U. La Malfa 153, I-90146 Palermo, Italy}

\author{C. Ferrigno}
\address{ISDC, Data Center for Astrophysics of the University of Geneva, chemin d'Ecogia, 16 1290 Versoix Switzerland}

\author{P. Esposito}
\address{INAF--IASF Milano, Via E. Bassini 15, I-20133 Milano, Italy}

\author{V. Mangano}
\address{INAF--IASF Palermo, Via U. La Malfa 153, I-90146 Palermo, Italy}

\begin{abstract}
Supergiant Fast X-ray transients are a subclass of high mass X-ray binaries displaying a peculiar and still poorly understood extreme 
variability in the X-ray domain. These sources undergo short sporadic outbursts ($L_{\rm X}$$\sim$10$^{36}$-10$^{37}$~erg s$^{-1}$), 
lasting few ks at the most, and spend a large fraction of their time in an intermediate luminosity state at 
about $L_{\rm X}$$\sim$10$^{33}$-10$^{34}$~erg s$^{-1}$. The sporadic and hardly predictable outbursts of supergiant fast X-ray transients were  
so far best discovered by large field of view (FOV) coded-mask instruments; their lower luminosity states require,  
instead, higher sensitivity focusing instruments to be studied in sufficient details.
In this contribution, we provide a summary of the current knowledge on Supergiant Fast X-ray Transients and explore 
the contribution that the new space mission concept LOFT, the Large Observatory For X-ray Timing, will be able to provide in the field of 
research of these objects.   
\end{abstract}


\end{frontmatter}

\parindent=0.5 cm

\section{Introduction: the Supergiant Fast X-ray Transients}
\label{sec:intro} 

Supergiant Fast X-ray transients (SFXTs) are a subclass of supergiant 
X-ray binaries \citep[SGXB, e.g.,][]{sguera05,negueruela06} comprising 
a neutron star (NS) accreting from the wind of its supergiant companion. 
At odds with the so-called ``classical'' SGXBs, which show a nearly persistent X-ray luminosity 
($\sim$10$^{35}$--10$^{37}$~erg s$^{-1}$), 
SFXT sources exhibit only sporadic outbursts reaching $\sim$10$^{37}$ erg s$^{-1}$ and 
lasting from minutes to hours. In-between these outbursts, SFXTs spend long intervals of 
time in lower luminosity states ($\sim$10$^{33}$-10$^{34}$~erg s$^{-1}$) during which 
a less pronounced flaring activity has also been observed \citep{bozzo10,sidoli10,bodaghee11}. 
These flares occur on time scales similar to that of the brightest outbursts (few thousands seconds), 
but the peak luminosity reaches $\sim$10$^{35}$~erg s$^{-1}$ at the most. 
The X-ray luminosity of the faintest emission state of SFXTs can be as low as 10$^{32}$~erg s$^{-1}$, i.e. 
a factor of $\sim 10^5$ lower than the brightest outbursts. 
To date, around 15 SFXTs have been identified. Five of these have extreme variability (luminosity dynamic range $\sim$10$^5$), 
whereas the others display a more moderate behavior (luminosity dynamic range $\sim$10$^3$) and are termed ``intermediate SFXTs'' 
\citep[for a recent list see, e.g.][and references therein]{falanga11}.  

Except for their peculiar X-ray variability, SFXT sources share many properties with previously known SGXBs. 
Measured orbital periods in SFXTs range from 3 to 50 days, and are thus similar to those of other SGXBs. 
The only exception is the SFXT IGR\,J11215-5952, with an orbital period of 165 days \citep{romano07,romano09}.  
The relatively high eccentricity inferred in two SFXTs ($\sim$0.3--0.7; {\small Zurita-Heras \& Chaty 2009, A\&A, 493, 1})
suggested that some of these systems might be characterized by somewhat more elongated orbits than classical SGXBs, but this 
conclusion seems not to be applicable to the most extreme SFXTs, which also have short orbital periods 
(IGR\,J16479-4514: P$_{orb}$=3.32~days \citep{jain09}; IGR\,J17544-2619: P$_{orb}$=4.92~days \citep{clark09}). In these cases, only a very reduced  
eccentricity (if any) can be assumed to avoid that the orbit of the NS gets too close to the surface of the supergiant companion.  
Spin periods have been securely measured only in three intermediate SFXTs and in IGR\,J11215-5952, and the 
estimated values ($\sim$100-1000~s) are compatible with those expected from classical SGXBs \citep[][]{bodaghee12}.  
None of the most extreme SFXTs have shown clear evidence for pulsations: the 4.7~s pulsations detected from 
IGR\,J18410-0535 were so far never confirmed \citep{bozzo11} and only a hint of detection was found for possible 
pulsations at 71~s from IGR\,J17544-2619 \citep{drave12}.  In two extreme SFXTs some evidence was found for 
spin periods as large as 1000-2000~s \citep{smith06}.  

The similarity between the geometrical and physical properties of the
SGXBs and SFXTs makes it difficult to develop a 
self-consistent scenario to interpret the behaviour of both systems in the X-ray domain.   
  
According to the ``extremely clumpy wind model'', the outbursts of SFXTs might be caused by the sporadic accretion of 
exceptionally dense clumps of material (populating the inhomogeneous wind of the supergiant star) onto the compact object     
\citep{zand05, walter07}. Early-type stars are indeed characterized by highly structured and variable massive winds 
which have an inherently clumpy nature \citep{oskinova07}. In this case, clumps with densities $\sim$10$^4$-10$^5$ higher than 
the surrounding stellar wind would be required to match the dynamic range of the X-ray luminosity in SFXTs. These clumps are expected to 
obscure (partially) the X-ray source when passing close to it, and thus in a few cases the measured variations of the local absorption 
column density during intense SFXT activity episodes were associated to these passages \citep{rampy09,bozzo11}. 
Recent simulations on clumpy wind accretion \citep[][; see in particular Fig.3 in the paper]{oskinova12} 
showed that this mechanism cannot explain the complete 
behavior of the SFXTs in X-rays. Indeed, beside the large X-ray luminosity variations produced by the accretion of these clumps 
(every few thousands seconds), the predicted long-term averaged X-ray luminosity for these objects (on orbital time-scales)  
would be close to that expected for classical SgHXBs and orders of magnitude higher than that measured from the SFXTs 
\citep{romano11}.  This is particularly puzzling for the short-orbital period SFXTs, for which also the possibility of having 
very large eccentric orbit to alleviate the problem seems to be ruled out for geometrical constraints.  

At least for the extreme short orbital period systems, other mechanisms might be required to explain their observational properties. 
Among these, it was proposed that centrifugal and magnetic ``gating'' mechanisms, due to the rotation of the NS  
and the intensity of its magnetic field, can halt (most of) the accretion flow during the orbital motion and 
reduce the average luminosity \citep{grebenev07,bozzo08}. In particular, the magnetic gate model can produce a wide dynamic range in X-ray luminosity 
($\gtrsim10^4$) even if accretion is taking place from a mildly inhomogeneous wind (small clumps) at the price of assuming that 
the NS is endowed with a long spin period ($>$1000~s) and a very strong magnetic field ($\gtrsim$10$^{14}$~G). 
As only very little is known about the SFXT spin periods and direct evidence of magnetic field in the required range (e.g. cyclotron lines) 
are difficult to obtain, the phenomenology of the SFXT remains still a matter of debate.

In the following, we show how the instruments on-board the new ESA space mission candidate LOFT, the Large Observatory For X-ray Timing, 
will be able to contribute to the SFXT science and significantly deepen our understanding of these objects.   

\section{LOFT, the Large Observatory For X-ray Timing} 
\label{sec:loft} 

LOFT \citep{feroci12}, the Large Observatory for X-ray Timing, is one of the four M3 space mission concepts 
selected by the European Space Agency (ESA) on February 2011. It is presently competing 
for a launch opportunity in 2022-2024. The mission is mainly devoted to the study of timing and spectral 
properties of X-ray emission in the proximity of black holes and neutron stars, in order to test general relativity in the 
strong field regime and constrain the neutron star equation of state \citep{feroci11}.   

The main instrument on-board LOFT, the Large Area Detector (LAD), is a non-imaging collimated experiment with an unprecedented large 
collecting area for X-ray photons (2-80~keV) reaching $\sim$10~m$^2$ at 8~keV. This instrument will provide a total of $\sim$240000~cts~s$^{-1}$ for 
a 1~Crab source and achieve a spectral resolution of $\sim$260~eV in the energy band 2-30 keV (in the 30-80~keV range only a 
coarse energy resolution will be available). Single-anode events will be endowed with an even better energy resolution ($<$200~eV), 
approaching the performance of CCD-based X-ray telescopes. The time resolution of the LAD is $<$10~$\mu$s.
This large area instrument could be conceived within the envelope of a medium-class size mission thanks to the 
technology of the large area Silicon drift detectors (SDDs) and the capillary plate collimators \citep[see][and references therein]{feroci11}. 

The LOFT payload comprises also a coded mask Wide Field Monitor \citep[WFM;][]{brandt12}. 
The WFM makes use of the same SDDs developed for the LAD, but with a design optimized for imaging purposes \citep{campana12}.  
It can achieve a timing ($\sim$10~$\mu$s) and spectral resolution ($\sim$300~eV) similar to that of the LAD, while observing more than 1/3 of the sky at once 
\citep[the main operating energy range is 2-50~keV; events in the 50-80~keV energy range will be mainly used to better evaluate the LAD background, see][]{feroci12}.  
An on-board intelligence, combined with a VHF transmitter, will be able to recognize and 
broadcast to the ground a number of bright impulsive events detected by the instrument within a delay of $<$30 s and a positional accuracy of 
$\sim$1~arcmin \citep[the so-called LOFT Burst Alert System, LBAS;][]{brandt12}. During observations of bright sources with the LAD, and in other cases 
in which the LAD will be occupying most of the available telemetry, the WFM will be operating with reduced resources, such that data might be available only with 
limited time ($\sim$300~s) and spectral (16 energy bands) resolution. 
The main science goal of the WFM is to detect new transient sources going off in the sky and monitor 
state changes of known celestial objects suitable for pointed observations with the LAD. However, given its unique capabilities, the instrument  
is able to carry out important scientific investigations by itself \citep[particularly relevant is the case of Gamma-Ray Bursts, X-ray bursts and 
transient outbursts such as those from SFXTs;][]{brandt12}.  

A summary of the capabilities of the WFM and the LAD is provided on the mission 
website\footnote{See http://www.isdc.unige.ch/loft/index.php/instruments-on-board-loft.}.

\section{LOFT observations of SFXTs}

In this section we provide an overview of the potential of LOFT to improve our knowledge on the 
outburst and lower activity X-ray emission of the SFXT sources. 

\subsection{Observations of SFXTs in outburst with the LOFT/WFM} 
\label{sec:wfmobs}

\begin{figure}[t]
\begin{center}
 \includegraphics[width=3.5in]{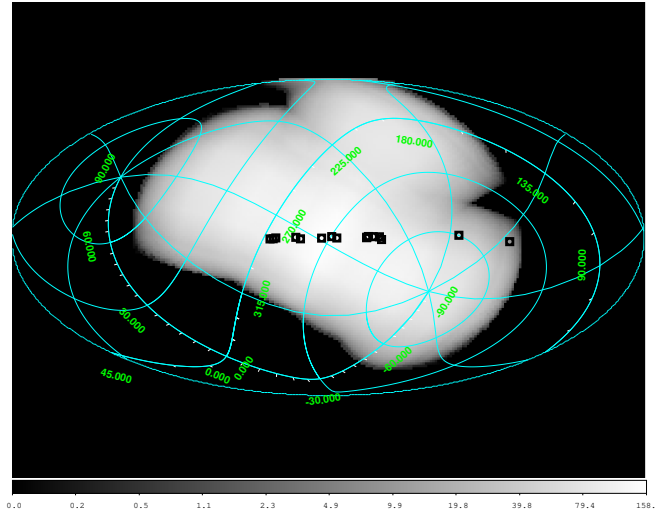}
 \caption{The WFM large FOV is able to catch all the known SFXTs at once while pointing, e.g., toward the Galactic center. 
 We plot for simplicity in grey the WFM FOV as provided by the LOFT WFM team (see text for details) 
 and superimposed the positions of all known 15 SFXTs (see Sect.~\ref{sec:intro}).}
   \label{fig:wfmfov}
\end{center}
\end{figure}

As described in Sect.~\ref{sec:intro}, the outbursts of almost all SFXTs have a sporadic nature and are thus best 
discovered with large FOVs. The unprecedented wide FOV of the LOFT/WFM is thus 
well suited to catch SFXT outbursts. In Fig.~\ref{fig:wfmfov}, we show the position of all 15 known SFXTs 
(see Sect.~\ref{sec:intro}) superimposed on the WFM FOV assuming the instrument is pointing toward the Galactic Center 
(the WFM FOV map is provided by the LOFT WFM team\footnote{See http://www.isdc.unige.ch/loft/index.php/preliminar-response-files-and-simulated-background.}). 
The WFM is able to catch all known SFXTs in a single shot while pointing toward the Galactic 
Center, thus giving the possibility to observe multiple outbursts from different sources simultaneously. 
In order to estimate the number of outbursts that can be expected from
the known SFXTs during the LOFT life-time, we searched into 
the Swift \citep{gehrels05} triggers for each source.
The BAT \citep{Barthelmy05} is scanning the sky regularly in a more uniform way with respect 
to other similar wide field instruments \citep[e.g., the INTEGRAL/ISGRI;][]{lebrun03,ubertini03}. 
We considered both BAT regular triggers, as disseminated through
GCN (independently on whether they were followed by a Swift slew to
point with the narrow-field instruments at them or not), and strong
on-board detections (those that would trigger the BAT with the current
triggering thresholds), as mostly reported in
\cite{romano09b,romano11b}
and at  http://www.ifc.inaf.it/sfxt/.
In particular, we estimated the expected number of outbursts from 
IGR~J11215$-$5952 which shows periodic flares at periastron every $\sim$165~d
\citep{romano07,romano09} as 2 per year. 
A summary of our findings is reported in Table~\ref{tab:summary}. 
As the LOFT mission is being planned to last for 
4~yrs plus 1~yr extension \citep{feroci12}, about 110 SFXT outbursts can be observed. This number is considered to be  
a conservative lower limit, as the WFM will have a higher sensitivity with respect to the Swift/BAT, and the much larger FOV is expected 
to contribute significantly to the discovery of new sources in this class. Note that a number of instrumental effects 
that will affect the final WFM performances (e.g., vignetting, off-axis angles, ...) are not included yet in the current simulations of SFXT observations 
with LOFT. These are indeed expected to be clarified only when the instrument properties will be fully consolidated and an advanced imaging simulator 
will be made available by the WFM team for general usage \citep{donnarumma12}.    
\begin{table}
\centering
\scriptsize
\caption{Expected number of outbursts observable from the known SFXT sources with the LOFT/WFM.} 
\begin{tabular}{@{}cccc@{}}
\vspace{0.5cm}\\
\hline
Source & Swift/BAT & \multicolumn{2}{c}{expected WFM detections} \\
\vspace{-0.2cm}\\
       & (outbursts/yr) & outbursts in 4~yr & outbursts in 5~yr \\
\hline
\vspace{-0.2cm}\\
 IGR J08408$-$4503    &   $>$2$^a$		&  8	&  10\\
 \vspace{-0.2cm}\\
 IGR J11215$-$5952    &    2$^b$		&  8	 & 10\\
  \vspace{-0.2cm}\\
 IGR J16479$-$4514    &    5$^c$	&	 20	&  25\\
  \vspace{-0.2cm}\\ 
 XTE J1739$-$302	    &    4$^c$		& 16	 & 20\\
  \vspace{-0.2cm}\\
 IGR J17544$-$2619   &     4$^c$		& 16	&  20\\
  \vspace{-0.2cm}\\
AX J1841.0$-$0536   &     2$^c$		&  8	&  10\\
 \vspace{-0.2cm}\\
 AX J1845.0$-$0433  &      $>$1$^a$	&	  4	&   5\\
  \vspace{-0.2cm}\\
 IGR J18483$-$0311  &      $>$1$^a$		&  4	&   5\\
 \vspace{-0.2cm}\\
 IGR J16328$-$4726   &     $>$0.5$^a$  &	  2	 &  2\\
 \vspace{-0.2cm}\\
 IGR J16418$-$4532    &    $>$0.5$^a$ & 	  2	&   2\\
 \vspace{-0.2cm}\\
 \hline
\multicolumn{4}{l}{$^{a}$: Estimated, lower limit, as number of BAT triggers yr$^{-1}$. See the reference}\\  
\multicolumn{4}{l}{section of http://www.ifc.inaf.it/sfxt/ for the complete reference list. }\\
\multicolumn{4}{l}{$^b$: This source shows periodic flares at periastron every $\sim$165~d}\\
\multicolumn{4}{l}{\citep{romano07,romano09}.}\\  
\multicolumn{4}{l}{$^c$: Number of on-board detections per year from \cite{romano09b,romano11b} }\\ 
\multicolumn{4}{l}{extrapolated to 4 and 5 years of LOFT mission.}\\  
 \end{tabular}
 \label{tab:summary} 
 \end{table}

\begin{figure}[t]
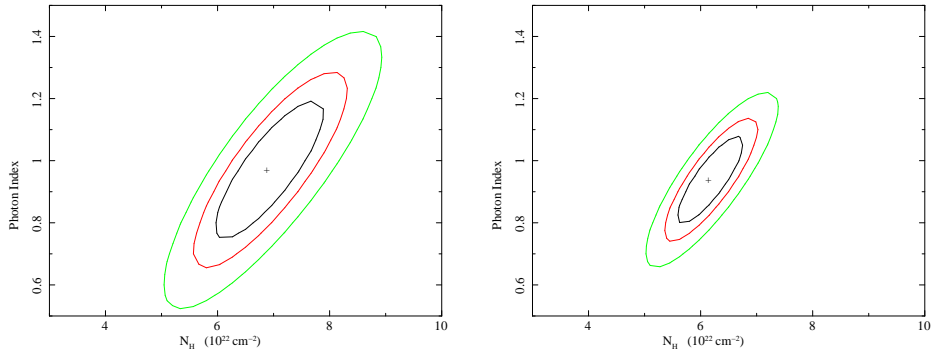

\begin{center}
\includegraphics[width=1.8in,angle=-90]{contours2ks.ps}
\includegraphics[width=1.8in,angle=-90]{contours5ks.ps}
\caption{Simulation of the observation of a bright outburst (2-10~keV flux of 5.9$\times$10$^{-9}$~erg cm$^{-2}$ s$^{-1}$ not corrected for absorption) 
from the SFXT IGR\,J16479-4514 with the LOFT/WFM. We show the 68\%, 90\% and 99\% c.l. contours for the uncertainties on the most relevant spectral 
parameters, i.e. the power-law photon index, $\Gamma$, and absorption column density, $N_{\rm H}$ (see text for details). 
Left figure is for an integration time of 2~ks, right figure for 5~ks. }
\label{fig:wfmspectrumoutburst}
\end{center}
\end{figure}

At odds with previously wide field monitor instruments that observed SFXT outbursts (i.e. Swift/BAT and INTEGRAL/ISGRI), the WFM on-board 
LOFT will have a significantly lower energy threshold (down to 2~keV), and will thus be able to catch the ``prompt emission'' of these events 
also in an energy range in which a number of spectral features are expected. Particularly interesting are 
changes in the local absorption column densities, which might be associated to the passage of clumps close to the accreting objects (see Sect.~\ref{sec:intro}).  
Variations of the power-law spectral index are also expected to provide indications of changes in the accretion flow geometry close to the 
surface of the NS \citep{bozzo10}. In Fig.~\ref{fig:wfmspectrumoutburst} we show how the LOFT/WFM 
would be able to see a bright outburst from the SFXT IGR\,J16479-4519\footnote{In this paper we use for the WFM and LAD spectral simulations the latest available 
response and background files of the instruments available on the official LOFT website (LAD response and background files v. 4.0; WFM response files v.2.0): http://www.isdc.unige.ch/loft.}. We assumed in this 
case a power-law spectral shape with photon index $\Gamma$=0.97, an absorption column density of 6.5$\times$10$^{22}$~cm$^{-2}$, and a cut-off 
energy of 13.5~keV as measured by \citet[][2-10 keV observed flux
5.9$\times$10$^{-9}$~erg cm$^{-2}$ s$^{-1}$]{romano08}. 
The contour plots show the uncertainties on the measured spectral parameters $\Gamma$ and $N_{\rm H}$ 
by assuming exposure times of 2~ks and 5~ks. The figure thus confirms that the LOFT/WFM is able to measure spectral parameters of the SFXT outbursts 
with a reasonably good accuracy and a time resolved spectral analysis can be carried out by using 
integration times as low as $\lesssim$2~ks. Measuring spectral variations in a large sample of events can provide crucial information on the triggering 
mechanisms of such phenomena \citep{bozzo11}.    

\begin{figure}[t]
\begin{center}
\includegraphics[width=2.2in,angle=-90]{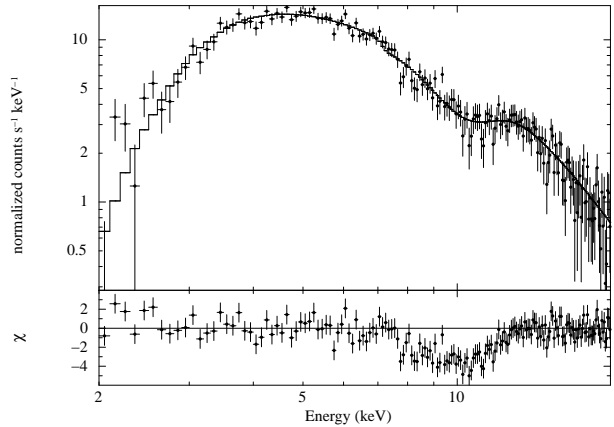}
\caption{Simulation of the observation of the same outburst in Fig.\ref{fig:wfmspectrumoutburst}, where a cyclotron absorption line at 10~keV 
was artificially added (see text for details). The residuals of the model that do not include the line are displayed in the bottom panel. 
The LOFT/WFM is well suited to discover these features with integration times as low as 2~ks.}
\label{fig:cyclotron}
\end{center}
\end{figure}

The broad-band high energy resolution spectra that the WFM will provide for the SFXT outbursts can also be used to search for those 
spectral features that might have gone undetected so far due to: (i) the limited energy resolution and the long integration times required for the currently available 
wide field instruments (INTEGRAL/ISGRI and Swift/BAT) to achieve the adequate signal-to-noise ratio (S/N), and (ii) the limited energy coverage of 
present focusing X-ray telescopes ($<$10~keV). As a speculative demonstration we show in Fig.~\ref{fig:cyclotron} a simulation in which we added to the 
spectrum of a bright outburst from IGR\,J16479-4514 (Fig.~\ref{fig:wfmspectrumoutburst}) a cyclotron line ({\sc gabs} in XSPEC) with a centroid energy 
of 10~keV \citep[we assumed for the depth $\tau$ and width $\sigma$ of the line parameters values typical of accreting NS, i.e. $\tau$=2 and 
$\sigma$=1.5~keV; see e.g.][and reference therein]{ferrigno11}. 
The figure shows that the LOFT/WFM is well suited to discover these features with integration times as low as 2~ks. 
The detection of cyclotron lines at energies $\gtrsim$10~keV, as found in many accreting NSs, gives a direct measurement 
of the magnetic field strength of the compact object, and might help distinguishing among some of the different models proposed to interpret the SFXT behavior 
(see Sect.\ref{sec:intro}). Indeed, a cyclotron line with a centroid energy in the range covered by the instruments on-board LOFT would indeed be indicative 
of relatively low magnetic fields ($\sim$10$^{12}$~G) compared to those expected in case of magnetars ($>$10$^{13}$-10$^{14}$~G). We note that, as the energy of the cyclotron 
line scales with the strenght of the magnetic field \citep[][and references therein]{ferrigno11}, this feature would fall outside the operating energy range of LOFT for a 
magnetar-like field. 

If not ruled-out by the detection of cyclotron lines at energies $\gtrsim$10~keV, evidence of magnetars in the SFXTs 
might still be inferred through alternative methods, as the measurement of the NS spin period and its derivative 
\cite[see, e.g.][]{finger10}. A description of the LOFT capabilities in searching for pulsations in the SFXT X-ray emission 
is provided in Sect.~\ref{sec:timing}.

\subsection{Observations of SFXTs in lower activity states with the LOFT/WFM} 
\label{sec:lowWFM}
   
Apart from the bright outbursts, SFXTs spend a large fraction of their life-time in lower emission states, with luminosities a factor 
10$^3$-10$^5$ lower than in outburst. A few SFXTs (4) were monitored by Swift for more than two years, and the low emission states could be classified and 
well characterized in terms of average luminosities and spectral parameters. We refer in the following to the classification presented by \citet[][see in particular 
Table~8 in their paper]{romano11}. For each of the emission states identified for the four objects, we reported in Table~\ref{tab:states} the exposure time needed 
to have a 5$\sigma$ detection with the LOFT/WFM. The high and medium emission states become accessible to the WFM in relatively short exposure times (a few ks to a few 
tens of ks). The instrument will thus be able to monitor sources in these states by providing several detections per day (almost all sources can be observed 
simultaneously during a single pointing, see Fig.~\ref{fig:wfmfov}). This will constitute an unprecedented database of historical flux variations in these systems, that might also 
help discover orbital periods in poorly known SFXT systems \citep[see, e.g.][and reference therein]{zurita09}. These  
measurements are also of crucial importance to understand the accretion mechanism in SFXTs, as they hold the potential to probe directly the accretion environment 
around compact objects and compare the latter with theoretical estimates of: (i) the distribution of clumps in the wind of supergiant stars, (ii) the structure and 
evolution of hot star winds. 

From Table~\ref{tab:states}  we note that the lowest emission states of the SFXT are hardly observable by the LOFT/WFM, and require long exposure times. 
Based on the 1~yr sky exposure map\footnote{http://www.isdc.unige.ch/loft/index.php/preliminar-response-files-and-simulated-background} expected at present, 
we estimated that the WFM will be able to reach a limiting flux of $\simeq$3$\times$10$^{-12}$~erg cm$^{-2}$ s$^{-1}$ for all the SFXTs (5$\sigma$ c.l.). 
For SFXT distances of 3-5~kpc \citep{rahoui08}, this corresponds to a luminosity of (3-10)$\times$10$^{32}$~erg s$^{-1}$, i.e. close to the lowest quiescent level displayed by these objects.    
\begin{table}
\centering
\scriptsize
\caption{Exposure time needed for 5$\sigma$ detections of some SFXTs during their different emission states (outside outbursts). The states 
have been defined according to Table~8 in \cite{romano11}.} 
\begin{tabular}{@{}cccc@{}}
\vspace{0.5cm}\\
\hline
Source Name &  \multicolumn{3}{c}{WFM exp. time in ks} \\
\vspace{-0.2cm}\\
            &  High State & Medium State & Low state \\
\hline
\vspace{-0.2cm}\\
 IGR J16479-4514   &   3.3		&  15.9	&  160.3\\
 \vspace{-0.2cm}\\
 IGR J1739-302    &   4.7		&  209.0	 &  --- \\
  \vspace{-0.2cm}\\
 IGR J17544-2619    &  42.3	&	410.0	&  --- \\
  \vspace{-0.2cm}\\
 IGR J18410-0535   &  12.3 & 58.6 & 540.0 \\
 \vspace{-0.2cm}\\
 \hline  
 \end{tabular}
 \label{tab:states} 
 \end{table}

\subsection{Observations of SFXTs with the LOFT/LAD} 

As the LAD is a non-imaging pointed instrument with a FOV of $\sim$1~deg., observing by chance a sporadic outburst 
from an SFXT requires too long exposure times. 
However, there exists at least one source in this class for 
which we know that outbursts always occur at the periastron passage, and are thus predictable. 
Indeed, the SFXT IGR\,J11215-5952 regularly displays outbursts every $\sim$165~d, and typically 
remains in a rather active X-ray phase for about 5~days around the event \citep{sidoli07}. 
Dedicated observational campaigns with the LAD can be planned to efficiently cover the beginning of the outburst, across 
the entire event, in order to monitor changes in the spectral parameters (e.g., the absorption column density) and 
possibly on the spin period ($\sim$187~s). The latter measurements can reveal the presence and intensity of accretion torques, 
in turn probing the geometry of the accretion flow around the NS and the properties of the environment from which 
the accretion takes place \citep[see e.g.][ and references therein]{bildsten97}. 

\begin{figure}[t]
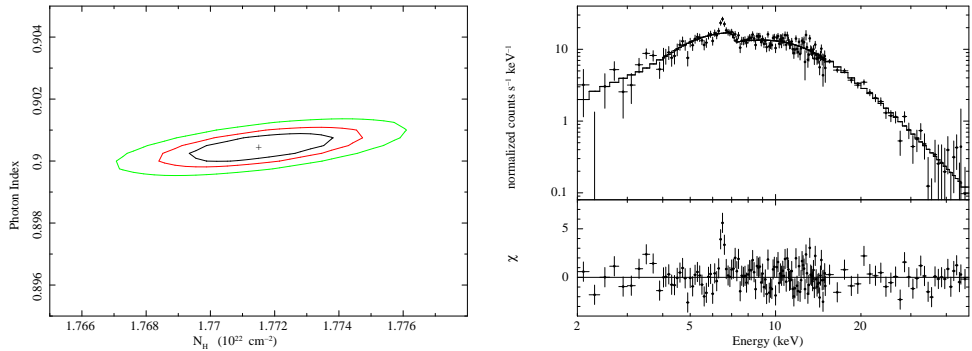

\begin{center}
\includegraphics[width=1.8in,angle=-90]{16479_lad.ps}
\includegraphics[width=1.8in,angle=-90]{16479_eclipse.ps}
\caption{{\it Left}: Simulation of an outburst from IGR\,J11215-5952 as observed with the LAD. In this case we reproduced the same spectral parameters 
measured by Swift/XRT in 2009 during 2.5~ks at the peak of the event \citep[observation ``001'' in][]{romano09b}. Here $\Gamma$=0.9, 
$N_{\rm H}$=1.77$\times$10$^{22}$~cm$^{-2}$, and the luminosity is 5$\times$10$^{35}$~erg s$^{-1}$. In 2.5~ks the LAD is able to collect about 20'000 cts from the source and recover the spectral parameters to within 1-2\% accuracy (compared to the 
30\% achieved by Swift/XRT). {\it Right}: Simulation of the eclipse ingress of the source IGR\,J16479-4514. The spectral parameters are the same as measured by 
XMM-Newton \citep{bozzo08b}, but the integration time is only 1~ks with the LAD. The iron line at 6.4~keV is clearly detected in the residuals from the fit 
shown in the bottom panel (the line was not included in the fit).}
\label{fig:ladsimul}
\end{center}
\end{figure}

In Fig.~\ref{fig:ladsimul}, we show a simulation of a typical outburst from the IGR\,J11215-5952 \citep[see Fig.~4 and 6 in][]{romano09b} with the LAD. 
For an exposure time of 2.5~ks, comparable to the integration time of some Swift/XRT pointings during the outburst of this source, 
the LAD is able to collect a total of 20000 counts, leading to a determination of the spectral parameters $\Gamma$ and $N_{\rm H}$ within 
a 1-2\% accuracy (compared to 30\% for Swift/XRT). This demonstrates that, if a similar event is observed with the LAD, 
significant variations of these spectral parameters (20-30\%) can be measured by using integration times as short as few tens of seconds.  
This will give the unique possibility to study the phenomenology of SFXTs on time scales shorter than the local dynamical times 
(few hundreds seconds), leading to new discovery windows for these objects so far completely unexplored.  

Note that the LAD will also be able to observe these sources not only during outburst, but also during fainter emission states 
(see Sect.~\ref{sec:intro}). Given the unprecedented large effective area and fine spectral resolution, the LAD is particularly well 
suited to detect and measure iron lines from the spectra of the SFXT sources in these states. These features can be used to study the physical properties 
of the accreting environment around compact objects, and in SFXT sources can provide evidence for the presence of massive clumps 
\citep[see e.g.][ and references therein]{bozzo08b,bozzo11}. 
In Fig.~\ref{fig:ladsimul}, we simulated the spectrum of the SFXT IGR\,J16479-4514 as was measured by XMM-Newton before the ingress to the eclipse. On that occasion 
the iron line emerged from the continuum spectrum due to the obscuration of the accreting NS by the supergiant companion. 
The integration time of the XMM-spectrum was 4~ks \citep[see Fig.~2 in][]{bozzo08b}. The LAD is able to detect the line at high significance 
with an exposure time of 1~ks. 

\subsection{Spin period measurements in SFXTs with LOFT} 
\label{sec:timing} 

The high time resolution and the large collecting area of the WFM and LAD   
(see Sect.~\ref{sec:loft}) are crucial to reveal pulsations in the X-ray emission of the SFXTs. 
The detection of relatively long spin periods ($\gtrsim$1000~s), possibly associated with strong period derivatives, might provide 
evidence for the presence of magnetars in these sources, and favor the ``gating'' accretion model 
(see discussion in \ref{sec:wfmobs}). So far, very little is known about 
the spin period of NSs in the SFXTs, as the active state in most of these sources last for just a few hours and makes the 
collection of enough counts to search for significant features in the corresponding power spectra challenging \citep[e.g.][and references therein]{bozzo11}. 
By assuming the currently envisaged performance of the instruments on-board LOFT (Sect~\ref{sec:loft}), we investigated the exposure time 
needed to measure a certain spin period with the LAD and the accuracy with which pulsations can be searched with the WFM during an  
outburst from an SFXT. We assumed a sinusoidal pulse profile and performed Fourier transform searches with 2$^{23}$-2$^{24}$ 
sampled frequencies. Under these assumptions, the pulsed fraction of the smallest detectable signal, $PF$, is related to the total number 
of counts available, $N_{ph}$, by $PF$$\propto$$N_{ph}$$^{-1/2}$. The results of our calculations are summarized in Table~\ref{tab:pulsations}.  
We considered three values of the source flux corresponding roughly to a bright outburst (6$\times$10$^{-9}$~erg/cm${^2}$/s), a fainter flare 
(6$\times$10$^{-10}$~erg/cm${^2}$/s), 
and a low emission state (10$^{-12}$~erg/cm${^2}$/s) of the SFXT prototype IGR\,J16479-4514 (see Sect.~\ref{sec:lowWFM}).  
\begin{table}
\centering
\scriptsize
\caption{Detection of pulsations from the X-ray emission of an SFXT in different emission states with 
the WFM and the LAD on-board LOFT \citep[see also][]{romano12}.} 
\begin{tabular}{@{}cccc@{}}
\vspace{0.5cm}\\
\hline
Instrument & Flux$^a$  & Pulsed Fraction\\
           & (erg/cm$^2$/s) & (\%) \\         
\hline 
\vspace{-0.2cm}\\
WFM$^b$  &  6$\times$10$^{-9}$ 		&  2.4 (5$\sigma$)\\
 \vspace{-0.2cm}\\
         &  6$\times$10$^{-10}$ 		&  66 (5$\sigma$), 56 (3$\sigma$)\\
\vspace{-0.2cm}\\    
LAD      &  6$\times$10$^{-9}$ 		&  0.1 (5~ks, 5$\sigma$)\\
\vspace{-0.2cm}\\    
         &  6$\times$10$^{-10}$ 		&  0.3 (5~ks, 5$\sigma$)\\
\vspace{-0.2cm}\\
         &  1$\times$10$^{-12}$ 		&  53.4 (10~ks, 3$\sigma$)\\
\vspace{-0.2cm}\\
 \hline    
\vspace{-0.2cm}\\ 
\multicolumn{3}{l}{$^a$: Flux is in the energy band 2-10~keV.}\\
\multicolumn{3}{l}{$^b$: An exposure of 5~ks is assumed.}\\
 \end{tabular}
 \label{tab:pulsations} 
 \end{table}
 
By using an integration time comparable to the typical duration of an SFXT outburst (i.e. a few ks), the WFM will be able to reveal  
pulsations in the X-ray emission recorded from the source during the event down to relatively small pulsed 
fractions (a few \%). Similar exposure times with the LAD permit to perform sensitive searches for pulsations also during the low 
emission states of these sources ($\sim$10$^{-12}$~erg/cm${^2}$/s).

\section{Conclusions} 

The SFXTs are characterized by a very unique and peculiar behavior in the X-ray domain, with triggering mechanisms still to be understood. 
The fast variability of these sources makes any observational campaign challenging and so far only wide field instruments with limited sensitivity 
and coverage at energies $<$15~keV were able to efficiently catch a high number of sporadic outbursts from these objects. 
Spectral features in the lower emission states (outside outbursts), which provide information on the accretion environment, could only be poorly 
investigated. This is mainly due to the large observational time required for the higher sensitivity focusing instruments to collect a sufficient number 
of photons and their limited energy coverage ($<$10~keV). 

The instruments on-board LOFT will be able to deepen our understanding of the SFXT sources. A large number of outbursts will be observed during the LOFT 
life-time by the wide FOV of the WFM, which will provide for these events also coverage of the prompt emission in the soft X-ray domain. Observations with the LAD 
during outbursts will make it possible to perform time-resolved
spectral analysis with integration times lower than the typical accretion 
dynamical time-scales close to the NS; this will open new discovery windows for the phenomena related to the SFXTs. The very large effective area of this instrument will 
also permit to detect spectral features in the X-ray emission of these sources that might help unveil the properties of their surrounding 
accretion environments. 

Both the WFM and the LAD will permit to search and measure pulsations in the X-ray emission from the SFXT sources with an unprecedented accuracy and down to very low pulsed fractions (few \%) 
and pulse periods up to several thousands seconds.

\section*{Acknowledgements}
EB acknowledges travel support from the Swiss Committee on Space Research (CSR) to attend the COSPAR 2012 meeting, 
and two anonymous referees for having contributed significantly to improve the manuscript. PR and VM acknowledge financial 
contribution from the contract ASI-INAF I/004/11/0. 


\end{document}